\documentclass[sigconf]{acmart}

\usepackage{verbatim}
\usepackage{multirow}
\usepackage{makecell}

\AtBeginDocument{%
  }

\copyrightyear{2023}
\acmYear{2023}
\setcopyright{rightsretained}
\acmConference[K-CAP '23]{Knowledge Capture Conference 2023}{December 5--7, 2023}{Pensacola, FL, USA}
\acmBooktitle{Knowledge Capture Conference 2023 (K-CAP '23), December 5--7, 2023, Pensacola, FL, USA}
\acmDOI{10.1145/3587259.3627571}
\acmISBN{979-8-4007-0141-2/23/12}

\begin{document}

\title{OLaLa: Ontology Matching with Large Language Models}

\author{Sven Hertling}
\email{sven@informatik.uni-mannheim.de}
\orcid{0000-0003-0333-5888}
\author{Heiko Paulheim}
\email{heiko@informatik.uni-mannheim.de}
\orcid{0000-0003-4386-8195}
\affiliation{%
  \institution{Data and Web Science Group, University of Mannheim}
  \city{Mannheim}
  \country{Germany}
}

\begin{abstract}

Ontology (and more generally: Knowledge Graph) Matching is a challenging task where information in natural language is one of the most important signals to process. With the rise of Large Language Models, it is possible to incorporate this knowledge in a better way into the matching pipeline.
A number of decisions still need to be taken, e.g., how to generate a prompt that is useful to the model, how information in the KG can be formulated in prompts, which Large Language Model to choose, how to provide existing correspondences to the model, how to generate candidates, etc.
In this paper, we present a prototype that explores these questions by applying zero-shot and few-shot prompting with multiple open Large Language Models to different tasks of the Ontology Alignment Evaluation Initiative (OAEI).
We show that with only a handful of examples and a well-designed prompt, it is possible to achieve results that are en par with supervised matching systems which use a much larger portion of the ground truth.
\end{abstract}



\maketitle

\section{Introduction}
From the first days of the Semantic Web and Linked Open Data, data integration has played a crucial role. Due to the open Web nature, everybody is able to create their own datasets and concept URIs without relying on a central instance. Thus everyone can create their own URI for the same real-world concept (a.k.a. non-unique name assumption). As a consequence, it is necessary to specify that two different URIs actually represent the same concept.
In Ontology or more general Knowledge Graph Matching, the task is to automatically find a set of correspondences between classes, properties, and instances of two different KGs such that the links are only generated if the corresponding concepts are equal.

In ontologies, the semantics are described with 1) natural language texts (e.g. \texttt{rdfs:label} or \texttt{rdfs:comment}) and 2)  relations to other concepts and formal axioms (e.g. taxonomies, domain and range definitions for properties). For a long time, the first was deemed to be only human interpretable, while the second was interpretable by humans and machines alike. Now with the arrival of large language models, this assumption is questionable, since computers are also able to process and interpret textual descriptions.

Thus, with the rise of transformer-based models~\cite{vaswani2017attention}, textual descriptions play an increasingly important role in Ontology Matching systems~\cite{he2022bertmap, hertling2022kermit, neutel2021towards}. However, there are still a lot of disadvantages in using those models.
The first one is the need for large training data. Most of the used language models are pre-trained and need a so-called \textit{head} (usually a simple dense layer at the very end) to be used in a classification setting. This neural network layer is initialized randomly and needs training to differentiate between matches and non-matches. This approach is usually referred to as \emph{fine tuning}.
Another disadvantage is the restricted amount of tokens (words/ pieces of text) that can be processed in such models. Thus the descriptions of concepts need to be short and precise.

With the development of Large Language Models (LLMs), it is possible to better capture the meaning of a text and also allow to reason about it.
One of the most famous models, ChatGPT\footnote{\url{https://chat.openai.com/}}, was developed by OpenAI and launched on November 30, 2022 to the public.
The interface (input and output) is purely based on texts which allows humans to have a chat with the bot. Due to its capabilities, it is applied in closely related fields, such as product matching~\cite{peeters2023using}.

There are also disadvantages for ChatGPT when applied to tasks such as KG matching.
The most important drawback is that it is not open source, but hidden behind an API. Thus, all achieved results are not reproducible (because OpenAI might change the model behind the API or even shut down a model that is afterwards not available anymore). Furthermore, it is not possible to have full access to the model, and thus no intermediate scores can be retrieved. Moreover, the company providing the closed-source models can charge the user with some cost per query. If the number of queries increases (e.g. with larger ontologies), it is questionable whether the use of ChatGPT is still economically sensible.
For those reasons, we will apply only open-source large language models to the task of Ontology Matching.

Applying LLMs for ontology matching requires a number of design decisions, including (1) the selection of models that actually perform best for this task, (2) how to present the matching task to the system, (3) how to generate candidates, (4) how to translate concepts into natural language text, (5) which prompts to use, and (6) detection of the final answer and extraction of confidences.
In this work, we provide a system that allows for systematic experimentation on all those questions. We show how to apply open-source LLMs to the task of ontology matching.

The contributions of this paper is as follows:
\begin{enumerate}
    \item implementation of different LLM-based matching components  in MELT\cite{hertling2019melt}
    \item evaluation of an LLM-based system against  in OAEI tracks
    \item analysis of the main driving factors for good results
\end{enumerate}

We show that with only a handful of examples for few-shot prompting and a well-designed prompt, it is possible to achieve results that are en par with supervised matching systems using a much larger portion of the ground truth.

The paper is structured as follows:
We briefly review related work in section~\ref{sec:related}. We present our approach coined \emph{OLaLa} in section~\ref{sec:approach}, followed by an evaluation, including an extensive ablation study, in section~\ref{sec:evaluation}. We conclude the paper with an outlook on future research.

\section{Related work}
\label{sec:related}

This section is divided into two parts. We first show approaches based on pre-trained language model which are related to the ontology matching task and afterwards we list related work based on large language models (both ChatGPT and open-source LLMs).
\subsection{Pretrained Language Models for Ontology Matching}
One of the first systems which applied transformer-based models to ontology matching was DITTO~\cite{li2020deep} in 2020.
They used BERT~\cite{devlin2018bert}, DistilBERT~\cite{sanh2019distilbert}, and RoBERTa~\cite{liu2019roberta} to detect if two entities are similar. One difference is that the schema is fixed (meaning that each entity has the same attributes).
They overcome the issue of small input sizes by reducing the amount of text with tf-idf weighting.
Neutel et al.~\cite{neutel2021towards} provides a system based on BERT but mainly for the automatic alignment of two occupation ontologies.
The BERTMap~\cite{he2022bertmap} system evaluates on datasets from the Ontology Alignment Evaluation Initiative (OAEI)~\cite{pour2023results}.
It includes a fine-tuning of the LMs and finally repairs the mapping in case of inconsistencies. The corresponding candidates are generated by sub-word inverted indices (which only include entity pairs that share many (sub-)words.
Our previous approach KERMIT~\cite{hertling2022kermit} is also fine-tuned either supervised (based on a fraction of the reference alignment) or unsupervised (based on a high precision matcher).
One difference to BERTMap is that the candidates are generated with Sentence-BERT~\cite{reimers2019sentence}. This embedding-based retrieval system can also include matching candidates that do not share any tokens (such as synonyms).

For ontology and KG integration, it is not only important to find equivalence relations between concepts and especially between classes but also other types of relations such as subsumption or meronymy relations. He et al.~\cite{he2023language} thus applied a language model to detect also the type of relation whereas \cite{hertling2023transformer} provides an already fine-tuned model based on various KGs such as DBpedia~\cite{auer2007dbpedia} and Wikidata. \cite{shi2023subsumption} used BERT models to predict subsumption relations in the e-commerce setting.

\subsection{Large Language Models for Ontology Matching}

Due to the fact that large language models (LLMs) are relatively new, only a few papers already exist.
We first present papers using ChatGPT: Peeters et al.~\cite{peeters2023using} use the chatbot to check if two product descriptions refer to the same product.
\cite{norouzi2023conversational} use ChatGPT for ontology alignment by providing the whole source and target ontology to the bot and asking for the final alignment between them.
They applied their approach to the conference track of OAEI (the ontologies are rather small) and achieved a high recall but the final F1 score is below the baseline (string equivalence) because of a low precision.
For ontology engineering, Mateiu et al.~\cite{mateiu2023ontology} tuned a GPT-3 model to translate between natural language text and OWL Functional Syntax. Thus it is used mainly to add axioms to an ontology and enrich it.
The closest related work is from Wang et al.~\cite{wang2023exploring}. They apply LLaMa 65B~\cite{touvron2023llama1}, GPT3.5, and GPT4 to the Biomedical Datasets for Equivalence and
Subsumption Matching~\cite{he2022machine}. The candidate generation is done by computing top k neighbors in an embedding space generated out of SapBERT~\cite{liu2020self} (a pre-trained BERT model designed for the biomedical domain). It is shown that especially GPT4 can outperform the state-of-the-art by a large margin.
Pan et al.~\cite{pan2023large} provide an overview of how LLMs can be used for Knowledge Graphs in general. Section 4.1.1 discusses the application of entity resolution and matching and section 4.3.3 ontology alignment. 

Most of the presented approaches use closed-source LLMs. This means that the results might not be reproducible after OpenAI discontinues some models or changes the models behind the API.
Thus we focus in this work on open-source models and present the system \emph{OLaLa}.

\section{Approach}
\label{sec:approach}
\begin{figure*}[ht]
  \centering
  \includegraphics[width=\textwidth]{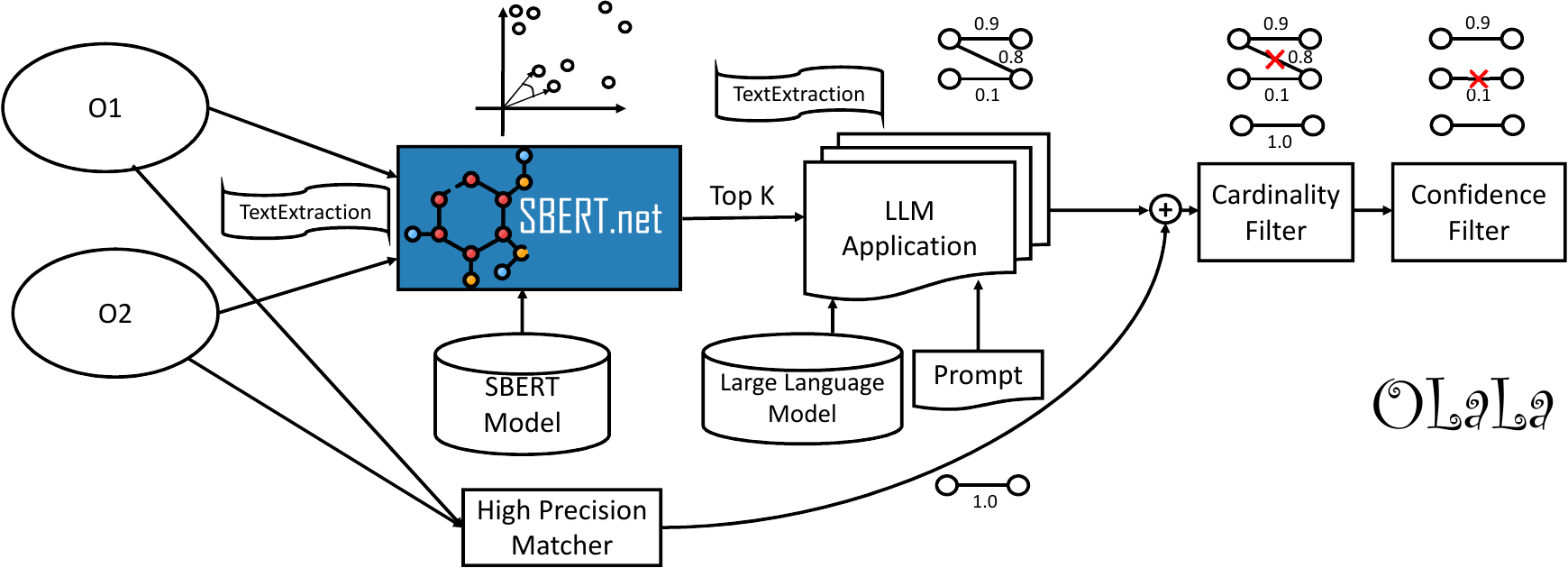}
  \caption{An overview of the \emph{OLaLa} system.}
  \label{fig:approach-overview}
\end{figure*}

Figure \ref{fig:approach-overview} shows an overview of the architecture of the \emph{OLaLa} system.
All components are implemented in MELT~\cite{hertling2019melt}, a framework for matcher development and evaluation. MELT is also used by the OAEI to package, submit, and evaluate the systems.
Thus, it is possible for the ontology matching community to reuse and customize each component in their own matching pipeline.
The implementation of \emph{OLaLa} is publicly available, and we provide a command line application\footnote{\url{https://github.com/dwslab/melt/tree/master/examples/llm-transformers}}
which allows to run the system and modify the most important parameters.

At the beginning, matching candidates need to be extracted from the two given input ontologies O1 and O2.
Afterwards, those candidates are included in the user-defined prompt and presented to the LLM. Two options are possible: 1) each correspondence is analyzed independently of each other 2) given a source entity, all possible target entities are presented and the LLM needs to decide which one is correct (or none of them).
The output of the high-precision matcher is added to ensure that the simple matches are included as well.
Finally, some filters are applied to fulfill the usual requirements for an alignment such as a 1:1 mapping (cardinality filter). The confidence filter at the end ensures that only correspondences with reasonably high confidence are returned. In the following sections, we will describe each step in more detail.

\subsection{Candidate Generation}
Due to the fact that the LLMs can usually not analyze the input ontologies as a whole (except small ontologies like those in the OAEI conference track, see \cite{norouzi2023conversational}), some correspondence candidates need to be generated. In this stage, only the recall is relevant and the higher the recall the better. Some of the related approaches apply an inverted index to find possible similar entities. This requires some textual overlap of those concepts. In \emph{OLaLa}, the well-known Sentence BERT models (SBERT) are used to generate those candidates. This allows a higher recall because it can also find similar entities without any textual overlap. The trained SBERT models are finetuned siamese BERT models on a huge set of paraphrases~\cite{reimers2019sentence}. SBERT as well as all LLMs only process text, but the input is an ontology. Thus it is necessary to verbalize the concepts into some natural language text. In MELT they are called \texttt{TextExtractors} (see section~\ref{subsec:approach_textextractors}).

For the candidate generation step, we use the so-called \texttt{Text\-Extractor\-Set}. It extracts all texts of a resource which are either labels (e.g. \texttt{rdfs:label}, \texttt{skos:prefLabel}, \texttt{schema:name}) or descriptions (e.g. \texttt{rdfs:comment}, \texttt{dc:description}, \texttt{schema:comment}). In addition to that, the URI fragment is extracted in case it does not contain more than 50\% digits. As a last step, all annotation properties are followed recursively and all labels of those resources are added as well.

All those extracted texts for each resource are embedded, and a semantic search is executed. It computes the cosine similarity between a list of query embeddings and a list of corpus embeddings and returns the top-k neighbors for each text. From those, we select the top-k best neighbors per resource. This procedure is repeated twice so that each of the input ontologies serves once as a query and one as a corpus.

\subsection{LLM Application}

There are two principal approaches how the candidates are presented to the LLM. The first one is \emph{binary decisions}, i.e., deciding whether one candidate is correct or not; the second is \emph{multiple choice} decisions, i.e., selecting the most likely correspondence for one concept from a set of possible targets.

\subsubsection{Binary Decisions}
Binary decisions are implemented in the class \texttt{LLMBinaryFilter}.
For each candidate correspondence, the source and target entity are verbalized as text and replaced in the prompt given by the user.
The output of generative models, such as the ones applied in this work, is always natural language text. To convert this into a binary decision, the following technique is applied: We search for target tokens/words that indicate the result (e.g. true/yes or false/no). If such a token is found, the generation process is directly stopped.
Due to the high computation cost, such an early stopping approach is useful to process a large number of candidates.
Up to now, only the decision is extracted and in case the model generates other texts like ``This is a correct match'',  we fail at detection.

To overcome this issue and also extract a specific confidence, we do the following.
If any of the target tokens is detected, then we retrieve the scores of the complete vocabulary and apply the softmax function to it. This corresponds to the probability that the word is generated at this position. We check the probability for all words in the positive class (e.g. yes, true) and take the maximum value which is normalized by the maximum value of the negative class (e.g. $\frac{0.4}{0.4 + 0.1}=0.8$ where $0.4$ corresponds to the probability of one token in the positive class like yes and $0.1$ corresponds to the maximum negative class tokens probability).
Thereby, we get a confidence between zero and one, and every confidence above 0.5 is a predicted positive token.

In case no positive or negative token is generated, the probabilities at the first generated token are used.
All those computations would not be possible with a model accessed by an API such as ChatGPT.\footnote{We also explored prompt engineering as an alternative to get to confidences, using prompts such as ``and also provide a confidence score with your answer``, but we observed that the LLM will often respond that it is not able to provide a specific confidence value, and even if it does, it is not easy to extract it out of the generated text. Therefore, we discarded that idea again.}

The default generation strategy\footnote{\url{https://huggingface.co/docs/transformers/main/en/generation_strategies}} is greedy such that each token with the highest probability is chosen and the generation process is continued with this text.
The implementation also allows to switch to e.g. contrastive search~\cite{su2022contrastive} but due to the usual short answers, it is neither necessary nor helpful.

\subsubsection{Multiple Choice Decisions}
Multiple choice decisions are implemented in the class \texttt{LLMChooseGivenEntityFilter}. It provides the LLM with more context such that for a given source entity all possible target entities with identifying letters are also shown. The task is to pick the one that represents the same entity or to generate a default answer such as ``none''. Confidences are extracted in the same ways as before. The normalization is applied to all possible outcomes including ``none``. There is also the possibility to use it directly for filtering such that the one with the highest confidence is kept and all others are removed. In case of a ``none`` prediction, all correspondences are removed.

\subsection{TextExtractors / Verbalizers}
\label{subsec:approach_textextractors}
In all the above cases, the extracted/verbalized texts for a given resource should be only one text and not multiple texts as for the candidate generation step.
Thus some of the possible extractors are now explained.

In addition to combining all texts from the \texttt{Text\-Extractor\-Set} explained before, an even simpler extractor called \texttt{Text\-Extractor\-Only\-Labels} is implemented. It extracts only one textual label which can originate from the following properties(in decreasing importance): \texttt{skos:\-pref\-Label}, \texttt{rdfs:\-label}, URI fragment, \texttt{skos:\-alt\-Label}, \texttt{skos:hidden\-Label}. This means if a \texttt{skos:\-pref\-Label} is detected, only this label is used.

Including more context in those examples is achieved by the \texttt{Text\-Extractor\-Verbalized\-RDF}. It selects all RDF triples from the corresponding KG where the resource is in the subject position. Those triples are verbalized - meaning that each subject, predicate, and object is replaced by the text of \texttt{Only\-Labels} extractor. All triples with a label-like property are skipped because the information is already included. As an example, the statement``:MA\_0000002 rdfs:subClassOf :MA\_0001112`` is converted to ``spinal cord grey matter sub class of grey matter``.

As a variation of the previous extractor, it is also tried out to provide the triples directly as serialized RDF. The default of the \texttt{Resource\-Description\-In\-RDF} extractor is to serialize to turtle format where the prefixes are used but the prefix definition is excluded from the generated text to make it shorter (other serializations can also be configured). If there are resources in the object position of the triples, they will be also replaced by a literal containing the corresponding label. 

\subsection{High-Precision Matcher}

The high-precision matcher is a simple matcher in MELT that efficiently searches for concepts with the exact same normalized label (or URI fragment if a label is not available).\footnote{\url{https://dwslab.github.io/melt/matcher-components/full-matcher-list}}
The normalization includes lowercasing, camel case, and deletion of non alpha-numeric characters. If there is only one such candidate for a concept, then it is matched.

\subsection{Postprocessing}
After the application of the LLM, the resulting alignment is further post-processed by filters.
To keep the matching pipeline simple, only two additional filters are applied.
The \emph{cardinality filter} ensures a one-to-one mapping which is usually required.
To solve the assignment problem, it is reduced to the maximum weight matching in a bipartite graph~\cite{cruz2009efficient} (class \texttt{Max\-Weight\-Bipartite\-Extractor} in MELT).

To further improve the alignment and remove correspondences that are likely to be incorrect, the confidence filter is applied.
All correspondences that do not have a higher or the same confidence as a predefined threshold value are excluded.

\section{Evaluation}
\label{sec:evaluation}

We evaluate our approach on the anatomy, biodiv, and commonkg tracks of OAEI\footnote{\url{http://oaei.ontologymatching.org/}}. Moreover, we show results on the Knowledge Graph track~\cite{hertling2020knowledge}, where only class correspondences are considered.
For all tracks, we compare \emph{OLaLa} against the three best-performing systems in the different OEAI tracks in the 2022 edition of the OAEI~\cite{pour2023results}.
The evaluation was performed using the MELT framework on a server running RedHat with 256 GB of RAM, 2x64 CPU cores (2.6 GHz), and 4 Nvidia A100 (40GB) graphics cards.

\subsection{Final Configuration}
\label{sec:final_config}
For the final configuration, a lot of parameters need to be fixed. 
The SBERT model for the candidate generation step is set to \texttt{\hyphenchar\font=`\- multi-qa-mpnet-base-dot-v1},\footnote{\url{https://huggingface.co/sentence-transformers/multi-qa-mpnet-base-dot-v1}} and the value k during the top-k neighbors search is set to five.
This gives a balance between the number of generated correspondences as well as the achieved recall. The \texttt{Text\-Extractor\-Set} is used to generate multiple text representations of the resource to run the search in the embedding space. 

The LLM model is set to \texttt{upstage/\-Llama-2-70b-instruct-v2}\footnote{\url{https://huggingface.co/upstage/Llama-2-70b-instruct-v2}} and to generate the text in prompt 7 (see table \ref{tab:ablation_prompt}), i.e., a few-shot prompt with three positive and negative examples each\footnote{The positive and negative examples are taken from the anatomy track and used across all tracks.}, \texttt{Text\-Extractor\-Only\-Labels} is used. With this prompt, the binary decision approach is automatically selected. 
For the text generation, the maximum number of tokens (\texttt{max\_\-new\_\-tokens}\footnote{\url{https://huggingface.co/docs/transformers/main/en/main_classes/text_generation\#transformers.GenerationConfig}}) is set to 10 but this number of tokens is usually not reached because a positive or negative word is detected before.
The next parameter which is fixed is the temperature. The lower the value,  the more deterministic the results are (the token with the highest probability is chosen as the predicted token). With increased temperature, the outputs are more randomized (resulting in more creative texts). We set the temperature to zero such that the results are reproducible. Other generation parameters are set to their default values.

The cardinality filter does not require any parameters, and the value of the confidence filter is set to 0.5.
With this setting, we filter out all correspondences where the LLM predicts a negative word (such as ``no`` or ``false``).
Thus we do not need to tune the confidence value and do not require any training alignment for it.

\begin{table}[t]
    \caption{Overall Results of the default configuration of \emph{OLaLa}, compared to the respective three best systems in different OAEI test cases}
    \resizebox{\linewidth}{!}{%
	\begin{tabular}{|l|l|l|l|l|l|l|}
		\hline
        \textbf{Test case}            & \textbf{System}    & Prec  & Rec & $F_1$ & Size & Time \\
        \hline \hline 
        \multicolumn{7}{|c|}{\textbf{Anatomy}}\\ \hline
		\multirow{5}{*}{mouse-human}  
                                & Matcha           & 0.951 & \textbf{0.930}  & \textbf{0.941} & 1482 & 0:00:37 \\\cline{2-7}
                                & SEBMatcher       & 0.945 & 0.874  & 0.908 & 1402 & 9:53:22 \\ \cline{2-7}
                                & \textbf{OLaLa}   & 0.914 & 0.891  & 0.902 & 1478 & 2:41:23 \\ \cline{2-7}
		                      & LogMapBio        & 0.873 & 0.919  & 0.895 & 1596 & 0:19:43 \\ \cline{2-7}
                                & String Baseline  & \textbf{0.997} & 0.622  & 0.766 & 946 & - \\ \cline{2-7}
        \hline
        \hline
        \multicolumn{7}{|c|}{\textbf{Common KG}}\\ \hline
        \multirow{5}{*}{\makecell[l]{nell-dbpedia}}     
                                                  & \textbf{OLaLa}   & \textbf{1.000} & \textbf{0.922}  & \textbf{0.960} & 120 & 0:06:34 \\ \cline{2-7}
		                                         & KGMatcher+       & \textbf{1.000} & 0.910  & 0.950 & 117 & 2:43:50 \\ \cline{2-7}
                                                  & Matcha           & \textbf{1.000} & 0.910  & 0.900 & 104 & 0:01:00 \\ \cline{2-7}
                                                  & ATMatcher        & \textbf{1.000} & 0.800  & 0.890 & 104 & 0:03:10 \\ \cline{2-7}
                                                  & String Baseline  & \textbf{1.000} & 0.600  & 0.750 & 78 & 0:00:37\\
        \hline
        \hline
        \multicolumn{7}{|c|}{\textbf{Knowledge Graph (only class matches)}}\\ \hline
        \multirow{5}{*}{\makecell[l]{marvel-\\cinematic-\\marvel}}     
                                                  & \textbf{OLaLa}   & \textbf{1.000} & \textbf{1.000}  & \textbf{1.000} & 11 & 0:17:40 \\ \cline{2-7}
		                                         & ATMatcher        & \textbf{1.000} & \textbf{1.000}  & \textbf{1.000} & 11 & 0:04:36 \\ \cline{2-7}
                                                  & LogMap           & \textbf{1.000} & \textbf{1.000}  & \textbf{1.000} & 10 & 0:32:40 \\ \cline{2-7}
                                                  & LSMatch          & \textbf{1.000} & \textbf{1.000}  & \textbf{1.000} & 8  & 1:46:01 \\ \cline{2-7}
                                                  & String Baseline  & \textbf{1.000} & 0.600  & 0.750 & 8  & 0:02:40\\
        \hline
        \multirow{5}{*}{\makecell[l]{memoryalpha-\\memorybeta}}
		                                         & ATMatcher        & 0.830 & \textbf{0.710}  & \textbf{0.770} & 39  & 0:03:23 \\ \cline{2-7}
                                                  & LogMap           & 0.880 & 0.500  & 0.640 & 21  & 0:05:09 \\ \cline{2-7}
                                                  & \textbf{OLaLa}   & \textbf{1.000} & 0.350  & 0.530 & 24  & 0:35:03 \\ \cline{2-7}
                                                  & LSMatch          & \textbf{1.000} & 0.290  & 0.440 & 26  & 0:57:37 \\ \cline{2-7}
                                                  & String Baseline  & \textbf{1.000} & 0.290  & 0.440 & 19  & 0:01:50 \\
        \hline
        \multirow{5}{*}{\makecell[l]{memoryalpha-\\stexpanded}}
		                                         & ATMatcher        & \textbf{1.000} & \textbf{0.770}  & \textbf{0.870} & 34  & 0:02:04 \\ \cline{2-7}
                                                  & \textbf{OLaLa}   & \textbf{1.000} & 0.540  & 0.700 & 28  & 0:29:41 \\ \cline{2-7}
                                                  & KGMatcher        & \textbf{1.000} & 0.540  & 0.700 & 29  & 0:25:42 \\ \cline{2-7}
                                                  & LSMatch          & \textbf{1.000} & 0.540  & 0.700 & 25  & 0:20:38 \\ \cline{2-7}
                                                  & String Baseline  & \textbf{1.000} & 0.460  & 0.630 & 19  & 0:01:11 \\ 
        \hline
        \multirow{5}{*}{\makecell[l]{starwars-\\swg}}
		                                         & LogMap           & \textbf{1.000} & \textbf{0.800}  & \textbf{0.890} & 12  & 0:07:44 \\ \cline{2-7}
                                                  & \textbf{OLaLa}   & \textbf{1.000} & 0.600  & 0.750 & 13  & 0:38:49 \\ \cline{2-7}
                                                  & ATMatcher        & \textbf{1.000} & 0.600  & 0.750 & 13  & 0:04:24 \\ \cline{2-7}
                                                  & LSMatch          & \textbf{1.000} & 0.600  & 0.750 & 19  & 0:38:50 \\ \cline{2-7}
                                                  & String Baseline  & \textbf{1.000} & 0.400  & 0.570 & 9   & 0:02:52 \\ 
        \hline
        \multirow{5}{*}{\makecell[l]{starwars-\\swtor}}
		                                         & ATMatcher        & \textbf{1.000} & \textbf{0.870}  & \textbf{0.930} & 31  & 0:04:20  \\ \cline{2-7}
                                                  & KGMatcher        & \textbf{1.000} & \textbf{0.870}  & \textbf{0.930} & 30  & 0:43:57 \\ \cline{2-7}
                                                  & String Baseline  & \textbf{1.000} & 0.800  & 0.890 & 27  & 0:02:51 \\  \cline{2-7}
                                                  & \textbf{OLaLa}   & 0.920 & 0.800  & 0.860 & 30  & 0:45:47 \\ \cline{2-7}
                                                  & LogMap           & \textbf{1.000} & 0.730  & 0.850 & 28  & 0:07:10 \\ \cline{2-7}
        \hline \hline
        \multicolumn{7}{|c|}{\textbf{Biodiv}}\\ \hline
        \multirow{4}{*}{\makecell[l]{envo-\\sweet}}
                                                  & LogMap           & 0.781 & \textbf{0.656}  & \textbf{0.713} & 676  & 0:00:25 \\ \cline{2-7}
                                                  & LogMapBio        & 0.753 & 0.652  & 0.699 & 697  & 1:00:03 \\ \cline{2-7}
                                                  & LogMapLt         & \textbf{0.829} & 0.594  & 0.692 & 576  & 0:07:32 \\ \cline{2-7}
                                                  & \textbf{OLaLa}   & 0.431 & 0.613  & 0.510 & 1145  & 6:55:19 \\ \cline{2-7}
        \hline

        \multirow{4}{*}{\makecell[l]{gemet-\\anaee}}
                                                  & AML (2021)       & \textbf{0.976} & 0.764  & \textbf{0.839} & 359  & 0:00:21 \\ \cline{2-7}
                                                  & ATMatcher (2021) & 0.631 & \textbf{0.919}  & 0.748 & 486  & 0:00:08 \\ \cline{2-7}
                                                  & \textbf{OLaLa}   & 0.565 & 0.916  & 0.699 & 542  & 4:28:07 \\ \cline{2-7}
                                                  & LogMapLt         & 0.840 & 0.458  & 0.593 & 182  & 0:00:03 \\ \cline{2-7}
        \hline
	\end{tabular}
 }
	\centering
	
	\label{tab:overall}
\end{table}

\subsection{Results and Discussion}
Table~\ref{tab:overall} shows the overall results of \emph{OLaLa} across the different tracks in the configuration above. 
Although it might be possible to tweak the parameters per track to achieve better results, we use only one configuration across all tracks in order to show a fair comparison.

We can see that in many test cases, \emph{OLaLa} scores among the top 3 systems, delivering good results with an out-of-the-box setup. It is worth mentioning that the other approaches often use domain-specific knowledge (especially in the biomedical domain) and/or extensively utilize the structure of the ontologies, while \emph{OLaLa} solely relies on the textual descriptions of entities.\footnote{For reasons of completeness, we should mention that we use three examples from the anatomy track for our few-shot prompt. Thus, one could argue that there is minimal information leakage for the anatomy track. However, given the alignment size, we consider this neglectable. Moreover, we could have used examples from other tracks for anatomy, but we wanted to keep the prompt constant across all tracks.}

At the same time, it can be observed that the runtimes utilizing LLMs are very often much higher than those for other models. This can be observed in particular in the Biodiv track, where the runtime of \emph{OLaLa} is often a few hours, compared to other systems which can solve the respective tasks in under a minute. 

\subsection{Ablation Study}
\label{subsec:evaluation_ablation}
In this section, we investigate the impact of the different parts and parameters of the system on the final result.
Due to the fact that all combinations on all tracks would drastically increase the number of experiments, we restrict ourselves to the anatomy track and only modify one component while keeping the rest of the system stable to the final configuration introduced in section \ref{sec:final_config}.

\begin{table}[t]
    \caption{Performance of zero-shot bi-encoders (SBERT models) on the anatomy track. The best recall per $k$ is highlighted with bold print. Time is measured in seconds.}
\resizebox{\linewidth}{!}{%
	\begin{tabular}{|r|l|r|r|r|r|r|}
		\hline
		\textbf{k}           & \textbf{Model}                 & Prec  & Rec            & $F_1$ & Size & Time\\ \hline
		\multirow{5}{*}{10} & multi-qa-mpnet-base-dot-v1    & 0.034 & \textbf{0.985}   & 0.066 & 43,786 & 8  \\
		                     & all-mpnet-base-v2             & 0.034 & 0.983          & 0.066 & 43,625 & 8  \\
		                     & multi-qa-distilbert-cos-v1    & 0.035 & \textbf{0.985} & 0.067 & 43,071 & 8  \\
                              & all-distilroberta-v1          & 0.034 & 0.981          & 0.066 & 43,567 & 8  \\
                              & all-MiniLM-L12-v2             & 0.034 & 0.983          & 0.066 & 43,399 & 8  \\ \hline
        \multirow{5}{*}{5} & multi-qa-mpnet-base-dot-v1     & 0.066 & \textbf{0.978}   & 0.124 & 22,338 & 8  \\
		                     & all-mpnet-base-v2             & 0.066 & 0.972          & 0.124 & 22,204 & 8  \\
		                     & multi-qa-distilbert-cos-v1    & 0.067 & 0.973          & 0.125 & 22,025 & 7  \\
                              & all-distilroberta-v1          & 0.066 & 0.968          & 0.123 & 22,366 & 6  \\
                              & all-MiniLM-L12-v2             & 0.066 & 0.974          & 0.124 & 22,229 & 6  \\ \hline
		\multirow{5}{*}{3} & multi-qa-mpnet-base-dot-v1     & 0.107 & 0.964            & 0.193 & 13,649 & 8  \\
		                     & all-mpnet-base-v2             & 0.108 & \textbf{0.966} & 0.194 & 13,543 & 7  \\
		                     & multi-qa-distilbert-cos-v1    & 0.108 & 0.963          & 0.194 & 13,553 & 6  \\
                              & all-distilroberta-v1          & 0.106 & 0.958          & 0.191 & 13,696 & 7  \\
                              & all-MiniLM-L12-v2             & 0.107 & 0.964          & 0.193 & 13,611 & 7  \\ \hline
		\multirow{5}{*}{1} & multi-qa-mpnet-base-dot-v1     & 0.306 & 0.931            & 0.461 & 4,612 & 8  \\
		                     & all-mpnet-base-v2             & 0.307 & \textbf{0.935} & 0.463 & 4,615 & 14  \\
		                     & multi-qa-distilbert-cos-v1    & 0.307 & \textbf{0.935} & 0.463 & 4,620 & 6  \\
                              & all-distilroberta-v1          & 0.292 & 0.904          & 0.442 & 4,692 & 6  \\
                              & all-MiniLM-L12-v2             & 0.306 & 0.933          & 0.461 & 4,620 & 7  \\ \hline
	\end{tabular}
 }
	\centering
	\label{tab:recall}
\end{table}

\subsubsection{Candidate generation}
In this stage, the SBERT model and corresponding k value for neighbor search need to be selected.
The available pretrained models are already evaluated on 14 datasets which checks the performance of the sentence embeddings as well as on six datasets for the performance of semantic search\footnote{\url{https://www.sbert.net/docs/pretrained_models.html}}. The best three models of each evaluation are selected to be tested on the anatomy track. All models are publicly available via the huggingface model hub. Table \ref{tab:recall} shows the results grouped by the value k.
On the one hand, with increasing k, the number of generated candidates gets also much higher and results in a large runtime of the following LLM model. On the other hand, all correspondences which are not found in this stage cannot be part of the final result. Thus, only the recall value and alignment size are important at this step. The results correlate with the performance on the semantic search datasets which is why the \texttt{multi-qa-mpnet-base-dot-v1} is selected (the top performing system on those 6 datasets).
The parameter k is set to five because recall could be increased by 1.2 (from k=3 to k=5), whereas changing from k=5 to k=10 only increases the recall marginally, but nearly doubles the amount of marginally candidates.

\begin{table}[t]
    \caption{Performance impact of using different LLM models on the anatomy track. }
\resizebox{\linewidth}{!}{%
	\begin{tabular}{|l|l|l|l|l|l|}
		\hline
		  \textbf{Model}          & Prec           & Rec             & $F_1$ & Size & Time \\ \hline
		  \makecell[l]{meta-llama/\\Llama-2-7b-hf}              & 0.932          & 0.640           & 0.759 & 1,041 & 7:50:33  \\ \hline
           \makecell[l]{meta-llama/\\Llama-2-13b-hf}             & 0.806          & 0.820           & 0.813 & 1,543 & 1:35:15  \\ \hline
           \makecell[l]{meta-llama/\\Llama-2-70b-hf}             & \textbf{0.946} & 0.860           & 0.901 & 1,378 & 6:45:13  \\ \hline
           \makecell[l]{meta-llama/\\Llama-2-70b-chat-hf}        & 0.663          & 0.801           & 0.725 & 1,832 & 3:55:57  \\ \hline
           \makecell[l]{jondurbin/\\airoboros-l2-70b-2.1}    & 0.804          & 0.877           & 0.839 & 1,654 & 4:00:12  \\  \hline
           \makecell[l]{upstage/\\Llama-2-70b-\\instruct-v2} & 0.914          & \textbf{0.891}  & \textbf{0.902} & 1,479 & 2:40:18  \\ \hline
	\end{tabular}
 }
	\centering
	
	\label{tab:ablation_model}
\end{table}

\subsubsection{LLM Model}
Table \ref{tab:ablation_model} shows the performance achieved with different LLM models.
The selection of the analyzed models is done with the help of the huggingface LLM leaderboard\footnote{\url{https://huggingface.co/spaces/HuggingFaceH4/open_llm_leaderboard}}. Many of those models are based on LLama2~\cite{touvron2023llama2} and fine-tuned on a specialized dataset.
As of 01/09/2023, model \texttt{jondurbin/airoboros-l2-70b-2.1} is the leading system whereas \texttt{upstage/Llama-2-70b-instruct-v2} is a general model which was also the leader of the board at the time of release.

It can be observed that the F-measure increases with the model size except for the chat variant of LLama2. The reason might be that prompt 7 is more designed for completion than a chat. Model \texttt{upstage/Llama-2-70b-instruct-v2} is selected due to a high F-measure as well as a low runtime.

For all models, the following parameters for loading the models are used: \texttt{device\_map} is set to `` auto``, \texttt{torch\_dtype} is set to ``float16``, and \texttt{load\_in\_8bit} is set to ``true``. With those settings, the memory footprint of the models is reduced such that the 7B and 13B variants fit on one A100 (40GB) GPU and the 70B variants on 2 GPUs of the same type. 

\begin{table}[t]
    \caption{Performance impact of using different text extraction strategies on the anatomy track.}
	\begin{tabular}{|l|l|l|l|l|l|}
		\hline
		  \textbf{Text Extractor} & Prec           & Rec             & $F_1$ & Size & Time\\ \hline
    OnlyLabels       & 0.914 & 0.891 & 0.902 & 1478 & 2:41:23 \\ \hline
    VerbalizedRDF    & 0.929 & 0.884 & 0.906 & 1443 & 3:57:46 \\ \hline
    DescriptionInRDF & \textbf{0.943} & \textbf{0.915} & \textbf{0.929} & 1471 & 9:02:24 \\ \hline
	\end{tabular}
	\centering
	
	\label{tab:ablation_textextractor}
\end{table}

\subsubsection{Text Extractors}

Table~\ref{tab:ablation_textextractor} shows the results if the text extractor is modified. The \texttt{Only\-Label} extractor is the worst in terms of F-Measure but it is also the fastest one (due to the small size of the input that needs to be processed). It is nice to see that the LLM can easily deal with RDF serializations (as produced by \texttt{Description\-In\-RDF} extractor) and achieve an even higher F-Measure than  SEBMatcher and close to Matcha. For the final configuration, the \texttt{Only\-Label} extractor is used to decrease the runtime even though other extractors could improve the final results.

The few-shot prompts also contain verbalizations of concepts. Those are created according to the selected text extractor. We also tested to keep the original prompt but achieved better results by using the same text extractor for example creation and testing.

\subsubsection{Prompts}
Table~\ref{tab:ablation_prompt} shows the prompts used. Prompts 0-4 are zero-shot, meaning that no examples were provided. Prompt one tests if additional context information (e.g. what are the topics of the ontologies) improves the results. Prompts 2, 3, and 4 further try to guide the model to answer with yes/no. Prompt 5 uses one positive and one negative correspondence whereas prompt 6 uses three positives and three negatives. With those added examples it is possible to reach the best precision but the overall best F-Measure is achieved by adding a description of the task at the very beginning (prompt 7). However, it is remarkable that the second best results are achieved with a simple zero-shot prompt (prompt 0).

Prompts 8 and 9 are multiple-choice decisions, which are observed to be inferior to single decision ones.

The runtimes vary drastically. The main reason is that for some prompts the target tokens (like yes/no etc.) are generated very late or not at all. In such cases, the text completion takes rather long (even though the maximum number of new tokens is set to 10). Overall 22,288 examples are classified whereas the multiple choice decisions only need to predict 6,035 examples. Multiple choice prompts can reduce the runtimes, but achieve less good results.

\begin{table}[t]
    \caption{Impact of the LLM and the different post processing pipelines on the anatomy track. HP represents the high-precision matcher.} 
\resizebox{\linewidth}{!}{%
	\begin{tabular}{|l|l|l|l|l|l|}
		\hline
		  \textbf{Postprocessing}          & Prec           & Rec             & $F_1$ & Size & Time \\ \hline
		  Candidates & 0.066 & \textbf{0.978} & 0.125 & 22,289 & 0:00:37 \\
            \hspace{0.2cm}+ Cardinality  & 0.385 & 0.693 & 0.495 & 2,731 & 0:00:37 \\
            \hspace{0.5cm}+ Confidence  & 0.387 & 0.693 & 0.497 & 2,715 & 0:00:37 \\
            \hspace{0.2cm}+ LLM + Cardinality & 0.591 & 0.919 & 0.719 & 2,357 & 2:37:51 \\
            \hspace{0.5cm} + Confidence & 0.911 & 0.889 & 0.900 & 1,480 & 2:37:51 \\
            \hspace{0.2cm}+ LLM + HP + Cardinality & 0.593 & 0.921 & 0.721 & 2,356 & 2:37:51 \\
            \hspace{0.5cm} + Confidence & \textbf{0.914} & 0.891 & \textbf{0.902} & 1,478 & 2:37:51 \\\hline
	\end{tabular}
 }
	\centering
	\label{tab:ablation_filter}
\end{table}

\subsubsection{Postprocessing}
In this section, the influence of the postprocessing is analyzed. Table~\ref{tab:ablation_filter} shows the results when only the candidate generation step is executed and when each filter is additionally added.
Without the LLM model, we achieve an F-Measure of 0.497 when the full filter chain is applied.

When using the LLM and the cardinality filter, the F-Measure is already increased to 0.719. Still, there are a lot of incorrect correspondences even though one entity is only mapped to a maximum of one other entity. Thus, the confidence filter is applied which lifts the F-Measure to 0.9. Adding the results of the high-precision matcher provides a slight increase in both precision and recall.

\begin{table*}[p]
    \caption{Examples of the prompts used and the results achieved on the anatomy track.}
	\begin{tabular}{|r|l|r|r|r|r|r|}
		\hline 
ID & prompt          & Prec           & Rec             & $F_1$ & Size & Time\\ \hline 
\multicolumn{7}{|c|}{Zero-Shot}\\ \hline
0 &   \makecell[l]{Classify if the following two concepts are the same.\textbackslash n \\
\texttt{\#\#\#} First concept:\textbackslash n\{left\}\textbackslash n\texttt{\#\#\#} Second concept:\textbackslash n\{right\}\textbackslash n\texttt{\#\#\#} Answer:
}& 0.853 & 0.866 & 0.861 & 1535 & 4:19:12  \\ \hline

1 &   \makecell[l]{
Classify if two concepts refer to the same real word entity. \\This is an ontology matching task between the anatomy of human and mouse.\textbackslash n \\
First concept:\{left\}\textbackslash n Second concept:\{right\}\textbackslash nAnswer:
} & 0.541 & 0.715 & 0.616 & 2002 & 2:18:22 \\ \hline

2 &   \makecell[l]{
Is \{left\} and \{right\} the same? The answer which can be yes or no is
} & 0.354 & 0.633 & 0.454 & 2709 & 11:46:12 \\ \hline
	
3 &   \makecell[l]{
The task is ontology matching. Given two concepts,\\the task is to classify if they are the same or not.\textbackslash n\\
The first concept is:\{left\}\textbackslash n The second concept is:\{right\}\textbackslash n\\The answer which can be yes or no is:
} & 0.754 & 0.830 & 0.791 & 1669 & 2:16:27 \\ \hline
 
4 &   \makecell[l]{
Given two concepts decide if they match or not.\textbackslash n\\
First concept:\{left\}\textbackslash n Second concept:\{right\}\textbackslash nAnswer(yes or no):
} & 0.520 & 0.704 & 0.598 & 2052 & 2:14:51 \\ \hline

\multicolumn{7}{|c|}{Few-Shot}\\ \hline

5 &   \makecell[l]{
\texttt{\#\#\#} Concept one: endocrine pancreas secretion\\\texttt{\#\#\#} Concept two: Pancreatic Endocrine Secretion \texttt{\#\#\#} Answer: yes\textbackslash n\\
\texttt{\#\#\#} Concept one: urinary bladder urothelium\\\texttt{\#\#\#} Concept two: Transitional Epithelium \texttt{\#\#\#} Answer: no\textbackslash n\\
\texttt{\#\#\#} Concept one: \{left\} \texttt{\#\#\#} Concept two: \{right\} \texttt{\#\#\#} Answer:
} & 0.751 & 0.762 & 0.757 & 1537 & 5:24:22 \\ \hline

6 & \makecell[l]{
\texttt{\#\#\#} Concept one: endocrine pancreas secretion\\\texttt{\#\#\#} Concept two: Pancreatic Endocrine Secretion \texttt{\#\#\#} Answer: yes\textbackslash n\\
\texttt{\#\#\#} Concept one: urinary bladder urothelium\\\texttt{\#\#\#} Concept two: Transitional Epithelium \texttt{\#\#\#} Answer: no\textbackslash n\\
\texttt{\#\#\#} Concept one: trigeminal V nerve ophthalmic division\\\texttt{\#\#\#} Concept two: Ophthalmic Nerve \texttt{\#\#\#} Answer: yes\textbackslash n\\
\texttt{\#\#\#} Concept one: foot digit 1 phalanx\\\texttt{\#\#\#} Concept two:  \texttt{\#\#\#} Answer: no\textbackslash n\\
\texttt{\#\#\#} Concept one: large intestine\\\texttt{\#\#\#} Concept two: Colon \texttt{\#\#\#} Answer: no\textbackslash n\\
\texttt{\#\#\#} Concept one: ocular refractive media\\\texttt{\#\#\#} Concept two: Refractile Media \texttt{\#\#\#} Answer: yes\textbackslash n\\
\texttt{\#\#\#} Concept one: \{left\} \texttt{\#\#\#} Concept two: \{right\} \texttt{\#\#\#} Answer:
}

& \textbf{0.979} & 0.687 & 0.807 & 1063 & 2:44:41 \\ \hline

7 &   \makecell[l]{
Classify if two descriptions refer to the same real world entity (ontology matching).\textbackslash n\\
\{prompt 6\}
} & 0.914 & \textbf{0.891} & \textbf{0.902} & 1478 & 2:37:51 \\ \hline

\multicolumn{7}{|c|}{Multiple Choice Decisions Zero-Shot}\\ \hline

8 &   \makecell[l]{
The task is ontology matching (find the description\\ which refer to the same real world entity).\\
Which of the following descriptions\\fits best to this description: \{left\}?\textbackslash n
\{candidates\}\\Answer with the corresponding letter or ``none`` if no description fits. Answer: 
} & 0.656 & 0.769 & 0.708 & 1778 & 1:06:04 \\ \hline

\multicolumn{7}{|c|}{Multiple Choice Decisions Few-Shot}\\ \hline
9 &   \makecell[l]{
The task is ontology matching (find the description\\ which refer to the same real world entity).\\
Which of the following descriptions\\fits best to this description: endocrine pancreas secretion?\textbackslash n\\
\textbackslash t a) Islet of Langerhans\textbackslash n
\textbackslash t b) Pancreatic Secretion\textbackslash n\\
\textbackslash t c) Pancreatic Endocrine Secretion\textbackslash n
\textbackslash t d) Delta Cell of the Pancreas\textbackslash n
\\Answer with the corresponding letter or ``none`` if no description fits. Answer: c\textbackslash n\\
Which of the following descriptions\\fits best to this description: \{left\}?\textbackslash n
\{candidates\}\\Answer with the corresponding letter or ``none`` if no description fits. Answer: 
} & 0.648 & 0.770 & 0.703 & 1802 & 0:43:47 \\ \hline

 \end{tabular}

	\centering	
	\label{tab:ablation_prompt}
\end{table*}

\section{Conclusion and Outlook}
\label{sec:conclusion}

In this paper, we presented \emph{OLaLa}, an ontology matching system that is built on top of open-source large language models. We have shown that using such a model, especially in a few-shot setting, can yield competitive results, even if only based on textual descriptions.

In our ablation study, we have observed that model and parameter combinations can have a strong impact on the overall results, and it is likely that there is no one-parameterization-fits-all solution, i.e., different parameter sets might deliver optimal results for different matching problems. Therefore, we plan to more closely examine the automatic parameterization of our system.

OLaLa provides an experimentation base for different variations, such as new prompts (prompt engineering), and also prompting techniques, like generating knowledge in the form of text that is used as additional information during classification~\cite{knowledgeprompting} or Chain-of-Thought prompting~\cite{chainofthought} that also allows to generate an explanation why two concepts are the same. In early experiments, we have observed that generating additional explanations for all candidates results in large runtimes (for anatomy, the expected runtime exceeds four days) but it could be useful to generate explanations for the final alignment which contains way less correspondences, or creating explanations on demand.

As already shown, the text extractors make a huge difference in terms of F-Measure. The RDF serialization works best but also generates a lot of tokens which could be reduced by selecting important properties to be included.

Finally, the system should be more scalable such that it can also be applied to large KGs with instance matching (which is technically possible, but with large runtimes).
This could be achieved, e.g., by using a fast high-precision matcher to first find easy matches, and applying the LLM model only to edge cases.

\begin{acks}
The authors acknowledge support by the state of Baden-Württemberg through bwHPC
and the German Research Foundation (DFG) through grant INST 35/1597-1 FUGG.
\end{acks}

\bibliographystyle{ACM-Reference-Format}
\balance
\bibliography{main}

\end{document}